# Influence de l'élasticité des lignes d'arbres, des roulements et du carter sur les vitesses critiques de rotation d'une transmission par engrenages

# Effect of elasticity of shafts, bearings and housing on the critical rotational speeds of a gearbox

**Emmanuel RIGAUD et Jean SABOT**

Laboratoire de Tribologie et Dynamique des Systèmes. UMR 5513. Ecole Centrale de Lyon.
36, avenue Guy de Collongue. B.P. 163. 69131 Ecully Cedex. FRANCE.

**Résumé :** On s'intéresse au comportement dynamique d'une transmission équipée d'un engrenage hélicoïdal. Les lignes d'arbres, les roulements et le carter sont discrétisés par éléments finis. Le couplage élastique entre les roues dentées est caractérisé par une matrice de raideur généralisée. Le calcul de la réponse vibratoire induite par l'erreur statique de transmission montre que les plus grandes surcharges dynamiques s'exerçant sur les dentures correspondent à une excitation en résonance des modes de denture. Les simulations numériques effectuées montrent qu'une prédiction réaliste des vitesses critiques de rotation doit prendre en compte toutes les composantes de la transmission.

**Abstract :** The dynamic behaviour of a gearbox fitted out with a helical gear pair is studied. Shafts, bearings and housing are discretised using the finite element method. The elastic coupling between the toothed wheels is characterised by a 12x12 stiffness matrix. The calculation of the vibration response induced by the static transmission error shows that the highest dynamic mesh forces correspond to a resonant excitation of modes which have a high mesh stiffness energy. The numerical simulations show that a realistic prediction of the critical rotational speeds should take account of all the components of the gearbox.

**Mots clés :** Dynamique des Engrenages, Interactions Dynamiques, Analyse Modale, Eléments Finis.

**Keywords :** Gear Dynamics, Dynamic Coupling, Modal Analysis, Finite Element Method

## INTRODUCTION

Lors de la conception d'une transmission par engrenages, il est important de savoir prédire quelles sont les vitesses de rotation qui vont donner naissance aux plus grandes surcharges dynamiques sur les dentures et, par conséquent, aux plus forts niveaux vibratoires et acoustiques. Pour une transmission fonctionnant à vitesse moyenne de rotation constante, les sources d'excitations vibratoires, et en particulier l'erreur de transmission sous charge, sont des fonctions périodiques résultant de la superposition de nombreuses composantes harmoniques [1, 2]. Le but de notre étude est de définir quels sont les modes propres de la transmission qui vont être excités en résonance par l'une des composantes harmoniques de l'erreur statique de transmission et, à partir de l'introduction d'une modélisation dynamique globale, de préciser l'influence de chacune des composantes mécaniques d'une transmission sur la localisation des vitesses critiques de rotation.

## 1- MODELISATION DE LA TRANSMISSION ETUDIEE

La transmission étudiée est une boîte d'inversion équipée d'un engrenage parallèle hélicoïdal 49/49 dents, discrétisée par la méthode des éléments finis (code de calcul ANSYS). Les lignes d'arbres de la transmission sont modélisées par des éléments de poutre. Chaque roue dentée est modélisée par des masses et des inerties rapportées sur la ligne d'arbres. La zone de contact entre les dentures est modélisée par une matrice de raideur généralisée symétrique qui couple les 6 degrés de liberté de la roue menée aux 6 degrés de liberté du pignon. Cette matrice est définie à partir des caractéristiques géométriques de l'engrenage et de la valeur moyenne de la raideur d'engrènement. Le moteur et la charge entre lesquels est intercalée la transmission sont modélisés par des éléments d'inertie en rotation. Ces inerties sont reliées aux arbres par l'intermédiaire de raideurs en torsion qui modélisent l'élasticité des accouplements flexibles. Nous avons introduit, pour chacun des 4 roulements identiques de la transmission, une raideur axiale ($10^8$ N/m), deux raideurs radiales ($10^9$ N/m) situées dans deux directions perpendiculaires l'une à l'autre, et deux raideurs ($10^6$ Nm/rad) couplant la rotation des sections libres des arbres au carter (raideur en rotation). Le carter en acier est un parallélépipède rectangle de 450*280 *160 mm d'épaisseur 10 mm. Il est modélisé à l'aide d'éléments de plaques et d'éléments volumiques utilisés pour représenter les boîtiers supportant les roulements. Le modèle élastique de la transmission étudiée possède environ 1000 éléments et 7000 degrés de liberté (voir figure 1).

## 2- CALCUL DE LA REPONSE DYNAMIQUE DE LA TRANSMISSION

Pour l'étude des vibrations forcées de la transmission, nous supposons que la seule excitation de la transmission est l'erreur statique de transmission sous charge [2]. En régime stationnaire, cette erreur induit une excitation dont les principales contributions sont la fluctuation de la raideur d'engrènement et les défauts de géométrie de l'engrenage [1, 3]. L'équation matricielle qui gouverne la réponse dynamique de la transmission est la suivante :

$$\mathbf{Mx} + \mathbf{Kx} + \mathbf{k(t)Dx} = \mathbf{F} + \mathbf{E(t)}$$

Dans cette équation, **M** et **K** sont les matrices classiques de masse et de raideur et **F** est le vecteur des forces extérieures. La matrice **D** est dérivée des caractéristiques de l'engrenage, **E** est un vecteur induit par l'erreur statique de transmission et **k(t)** correspond à la fluctuation de la raideur d'engrènement. Dans la base modale définie à partir des caractéristiques mécaniques moyennes et en introduisant un amortissement visqueux équivalent, l'équation devient :

$$\mathbf{mq + cq + kq + g(t).d.q = s}$$

**m**, **c** et **k** sont les matrices diagonales de masse, d'amortissement et de raideur. **s** est le vecteur d'effort modal et **g(t).d** est une matrice non diagonale qui est induite par la fluctuation de la raideur d'engrènement et qui couple les équations du mouvement. Pour résoudre le système mis en place, on utilise la Méthode Spectrale Itérative [4] qui permet de calculer le spectre de la réponse en chaque degré de liberté du système.

Nous avons choisi une erreur statique de transmission sous charge résultant d'une variation harmonique de la raideur d'engrènement. La fréquence d'excitation est égale à la fréquence d'engrènement. La valeur moyenne de la raideur d'engrènement égale à $4.10^8$ N/m et l'amplitude crête de sa fluctuation harmonique égale à $4.10^7$ N/m. Les simulations numériques sont effectuées en régime stationnaire et pour un couple moteur constant égal à 500 Nm. Nous avons choisi un taux d'amortissement visqueux équivalent égal à 3% pour chaque mode propre.

## 3- MODES DE DENTURE

Parmi tous les modes propres, certains d'entre eux ont une forme propre qui présente un ventre d'amplitude dans la zone de contact entre les dentures. On peut identifier ces modes à partir de l'approche énergétique suivante. Pour chaque mode, on définit l'énergie de déformation $\mathbf{U_d}$ associée à la raideur d'engrènement, l'énergie de déformation totale $\mathbf{U_T}$ et leur quotient $\rho$.

$$U_d = 1/2 \,.( {}^t\Phi \, [k_d] \, \Phi )$$

$$U_T = 1/2 \,.( {}^t\Phi \, [k_T] \, \Phi )$$

et $$\rho = U_d / U_T$$

Les modes qui présentent le taux d'énergie $\rho$ le plus grand sont appelés **modes de denture**. Leur excitation donne lieu à de fortes amplifications de la charge s'exerçant sur les dentures.

## 4- INFLUENCE DE LA FLEXION DES ARBRES

Nous avons calculé les modes propres de la transmission pour un modèle n'incluant que les déformations de torsion des lignes d'arbres. L'analyse des formes propres des modes nous a permis de détecter deux modes de denture (**1531 Hz** et **3140 Hz**). Nous avons ensuite calculé les modes propres de la transmission en utilisant un modèle incluant les déformations de traction-compression et de flexion des lignes d'arbres. Le carter et les roulements étaient supposés rigides. On note la présence de trois modes de denture (**1251 Hz**, **3056 Hz** et **3682 Hz**). Contrairement aux autres modes propres de la structure, les formes propres des modes de denture présentent simultanément des déformations de flexion et des déformations de torsion. Ce phénomène s'explique par le fait que l'engrenage couple l'ensemble des degrés de liberté de la roue menée à l'ensemble des degrés de liberté du pignon.

La figure 2 montre l'évolution de la valeur efficace de l'effort de denture en fonction de la fréquence d'engrènement. On peut constater que les résonances des modes de denture génèrent les plus fortes charges sur la denture. L'écart entre le modèle de torsion-flexion et le modèle de torsion pure traduit un effet de la flexion des arbres. Il en résulte qu'un modèle de torsion pure ne permet pas en général de prédire correctement les vitesses critiques de rotation. On note de plus que ce type de modèle surestime le niveau des surcharges dynamiques.

## 5- INFLUENCE DES ROULEMENTS

Les résultats numériques présentées ci-dessous ont été obtenus à partir d'un modèle prenant en compte les déformations élastiques des roulements. La transmission présente de nouveaux modes de denture principaux (**1191 Hz**, **3365 Hz** et **4157 Hz**). Ces modes ne sont pas les premiers modes propres du système. L'évolution de la fréquence de ces modes par rapport à ceux associés au modèle avec roulements rigides montre que l'élasticité des roulements modifie les vitesses critiques de rotation.

### Influence de la raideur radiale

Nous avons fait varier la raideur radiale dans une plage comprise entre $10^8$ N/m à $10^{10}$ N/m. Cette plage correspond aux valeurs habituelles rencontrées dans les transmissions par engrenage. A titre comparatif, la valeur de la raideur moyenne d'engrènement de la transmission étudiée est de 4 $10^8$ N/m. La figure 3 présente les courbes d'égal niveau de la valeur efficace de l'effort de denture dans un plan raideur radiale des roulements/fréquence d'engrènement. Pour la boîte d'inversion étudiée et dans la plage de variation de raideur considérée ($10^8$N/m à $10^{10}$N/m), on observe les caractéristiques suivantes :

**1-** Les modes de denture principaux ont une fréquence propre qui augmente avec la raideur radiale des roulements. Pour les premiers modes de denture (fréquence propre inférieure à 2000 Hz), cette évolution est très marquée entre $10^8$ et $10^9$N/m, puis moins marquée à partir de $10^9$N/m. Pour les modes de denture situés à des fréquences plus élevées (fréquence propre supérieure à 2000 Hz), cette évolution est d'abord assez faible entre $10^8$ et $10^9$N/m, puis plus marquée à partir de $10^9$N/m.

**2-** L'évolution de la fréquence des modes de denture principaux s'accompagne d'une modification de la forme propre associée à ces modes si bien que les modes de denture principaux deviennent moins énergétiques.

**3-** Les modes de denture principaux (**1676 Hz**, **3349 Hz** et **4139 Hz** pour $K=10^8$ N/m) finissent par disparaître au profit de l'apparition de nouveaux modes de denture situés à des fréquences plus basses (respectivement **1364 Hz**, **3311 Hz** et **4053 Hz** pour $K=10^{10}$ N/m). Ce dernier comportement se traduit par le fait que les vitesses critiques de rotation sont plus faibles lorsque les raideurs radiales sont grandes.

**Influence de la raideur en rotation**

Nous avons aussi fait varier la raideur en rotation des roulements entre $10^4$ Nm/rad et $10^8$ Nm/rad. Cette plage permet d'étudier l'influence de divers types de roulements : roulements à billes (raideur en rotation faible), roulements à rouleaux cylindriques (raideur en rotation moyenne) et roulements à rouleaux coniques (raideur en rotation forte).

La figure 4 présente les courbes d'égal niveau de la valeur efficace de l'effort de denture dans le plan raideur en rotation des roulements/fréquence d'engrènement. Les fréquences propres des modes de denture principaux évoluent avec la raideur en rotation des roulements. Cette évolution est très marquée entre $10^6$ et $10^7$ Nm/rad. Les modes de denture principaux (respectivement **1110 Hz**, **3100 Hz** et **3914 Hz** pour K=$10^4$ Nm/rad) disparaissent au profit de l'apparition de nouveaux modes (respectivement **1259 Hz** et **3459 Hz** pour K=$10^8$ Nm/rad).

## 6- INFLUENCE DU CARTER

Afin d'évaluer l'influence des propriétés mécaniques (élasticité et inertie) du carter sur les vitesses critiques de rotation de la transmission, nous avons réalisé une analyse modale de la transmission complète. Le modèle prend en compte l'ensemble des composantes de la transmission c'est à dire l'engrenage, les arbres, les roulements, le carter et l'environnement. L'examen détaillé des nouveaux modes propres conduit à un certain nombre de remarques. (1) Dans la plage de fréquences considérée, les modes de denture sont nombreux mais l'énergie élastique (associée à la raideur d'engrènement) de chacun est assez faible. (2) Les déformations des lignes d'arbres sont couplées aux déformations du carter. Le comportement dynamique du carter interagit donc avec l'engrenage. (3) En raison des déformations du carter, la fréquence et la forme propre de chaque mode de denture sont différentes de celles calculées avec un carter rigide.

Nous avons comparé l'évolution de la valeur efficace de l'effort de denture avec les résultats précédents (carter infiniment rigide). Cette comparaison est présentée pour des couples raideur radiale-raideur en rotation des roulements respectivement égaux à $10^8$ N/m-$10^4$ Nm/rad (figure 5a) et $10^9$ N/m-$10^6$ Nm/rad (figures 5b).

Comme l'illustre la figure 5a, les propriétés mécaniques du carter ne modifient pas les modes de denture pour des raideurs radiales de $10^8$ N/m et des raideurs en rotation des roulements de $10^4$ Nm/rad. En effet, les déformations du carter ne sont pas couplées à celles des lignes d'arbres. Par contre, pour des raideurs de roulements plus élevées, le comportement dynamique de la transmission est très affecté par les propriétés élastiques du carter. Comme l'illustrent la figure 5b, les résonances sont plus nombreuses. Les fréquences de ces résonances ont sensiblement évolué et le niveau maximum de la surcharge dynamique maximale devient plus faible. Des études paramétriques nous ont montré que, pour la transmission étudiée, les raideurs en rotation jouent un rôle plus important que les raideurs radiales dans le couplage entre le carter et l'engrenage.

## CONCLUSION

Pour étudier le comportement dynamique d'une transmission équipée d'un engrenage hélicoïdal, nous avons modélisé l'ensemble des composantes de cette transmission par la méthode des éléments finis. Nous avons défini une méthode permettant, à partir d'une analyse modale de la transmission, d'identifier les modes de denture. Pour des vitesses de rotation du moteur comprises entre 0 et 3000 trs/min, ces modes, situés entre 0 et 4500 Hz, peuvent être excités par la fréquence d'engrènement ou, éventuellement, l'un de ses harmoniques. L'erreur statique de transmission génère alors les plus grandes surcharges dynamiques sur les dentures et, par conséquent, les plus forts niveaux vibratoires du carter.

Alors que les vitesses critiques de rotation sont habituellement définies à partir d'une étude du comportement des lignes d'arbres en torsion pure, nous avons mis en évidence l'influence de la flexion des lignes d'arbres, de l'élasticité des roulements et des propriétés mécaniques du carter sur ces vitesses critiques de rotation et sur les niveaux des surcharges dynamiques sur les dentures. Nous pouvons donc conclure que seule une modélisation globale, incluant les lignes d'arbres, les roulements et le carter peut permettre de prédire de façon réaliste les modes de denture et les vitesses critiques de rotation d'une transmission par engrenages.

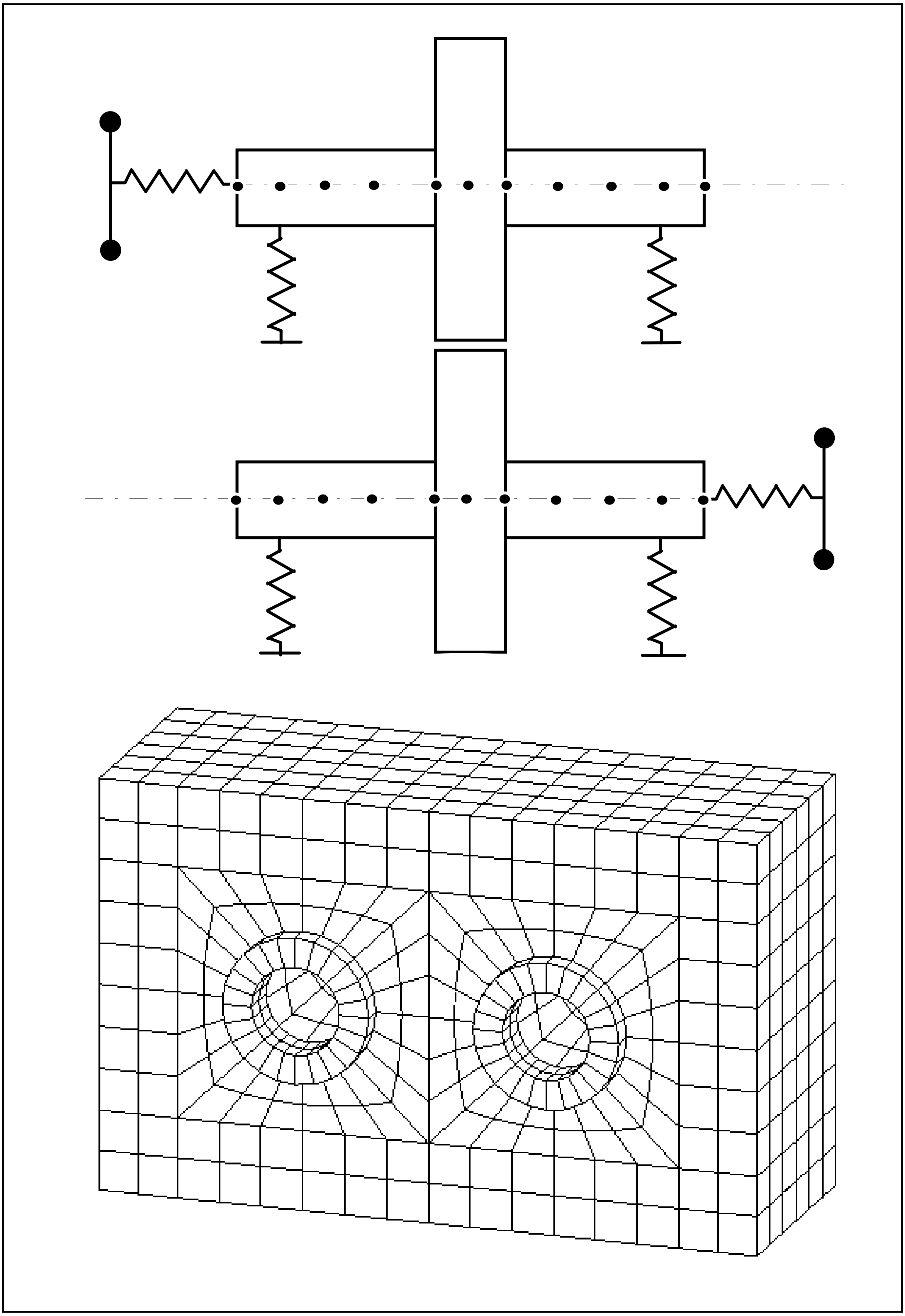

Figure 1. Modélisation des lignes d'arbres et du carter.

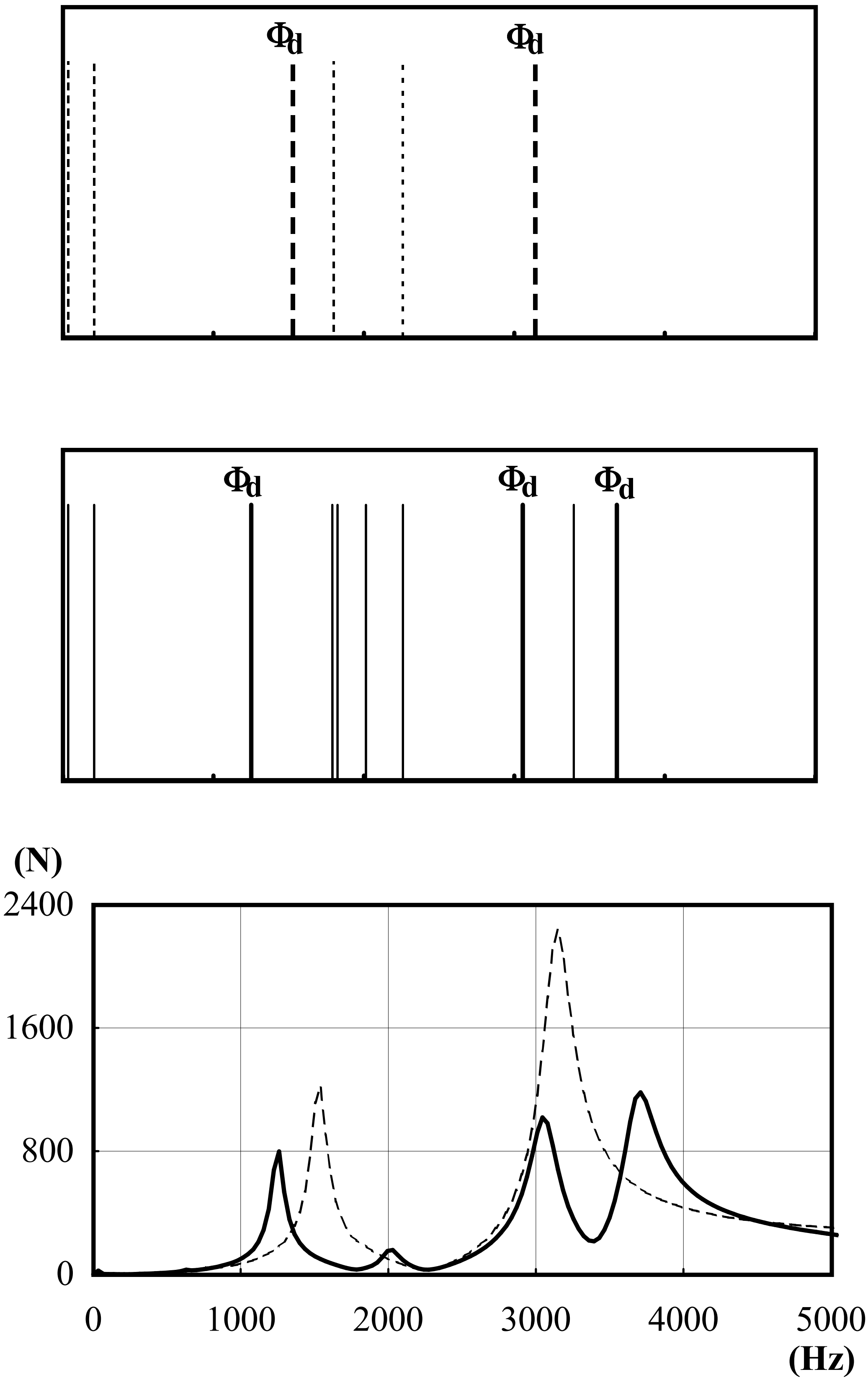

Figure 2. Evolution de la valeur efficace de l'effort de denture avec la fréquence d'engrènement.

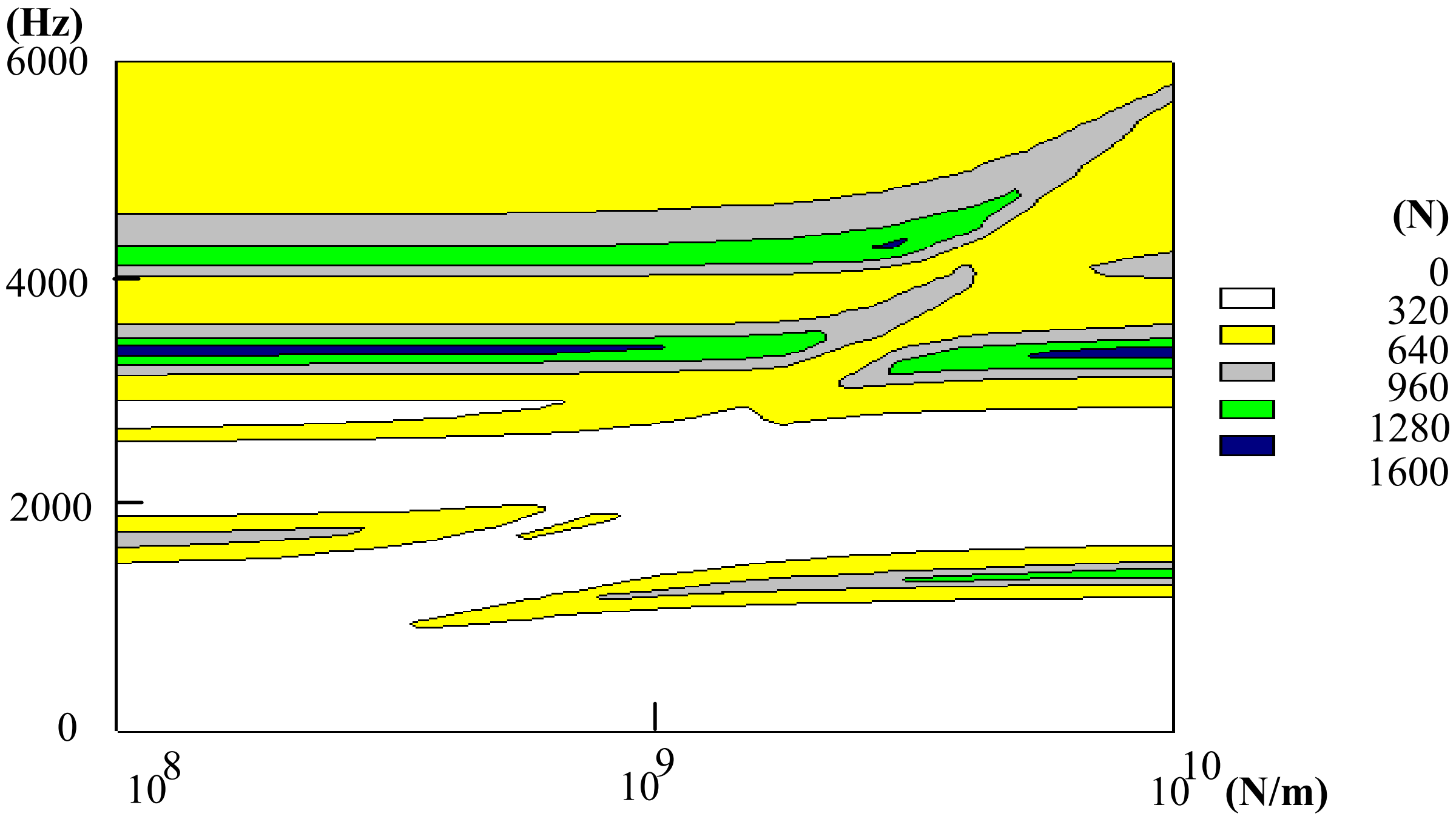

Figure 3. Evolution de la valeur efficace de l'effort de denture avec la raideur radiale des roulements (en abscisse) et la fréquence d'engrènement (en ordonnée).

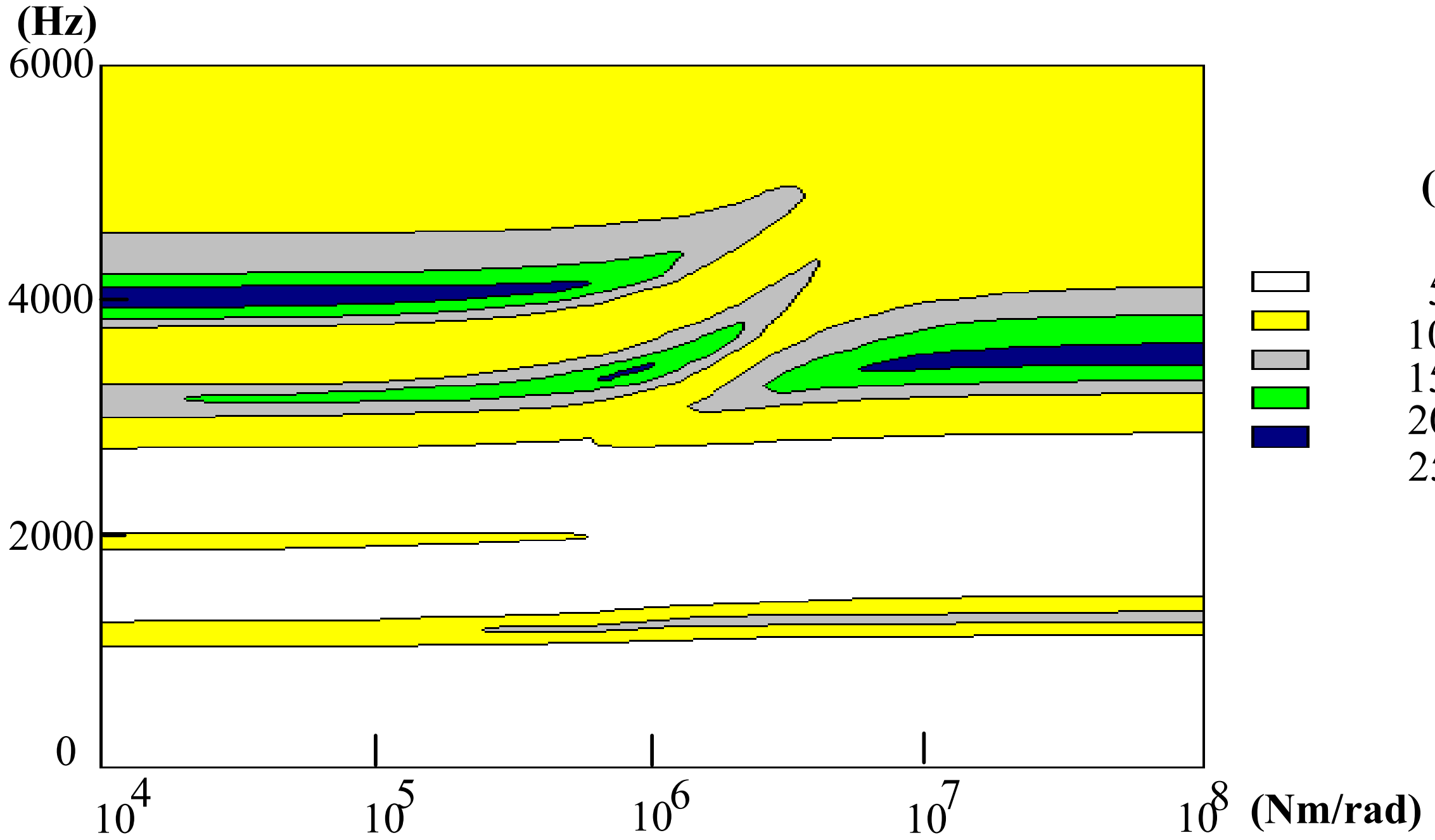

Figure 4. Evolution de la valeur efficace de l'effort de denture avec la raideur en rotation des roulements (en abscisse) et la fréquence d'engrènement (en ordonnée).

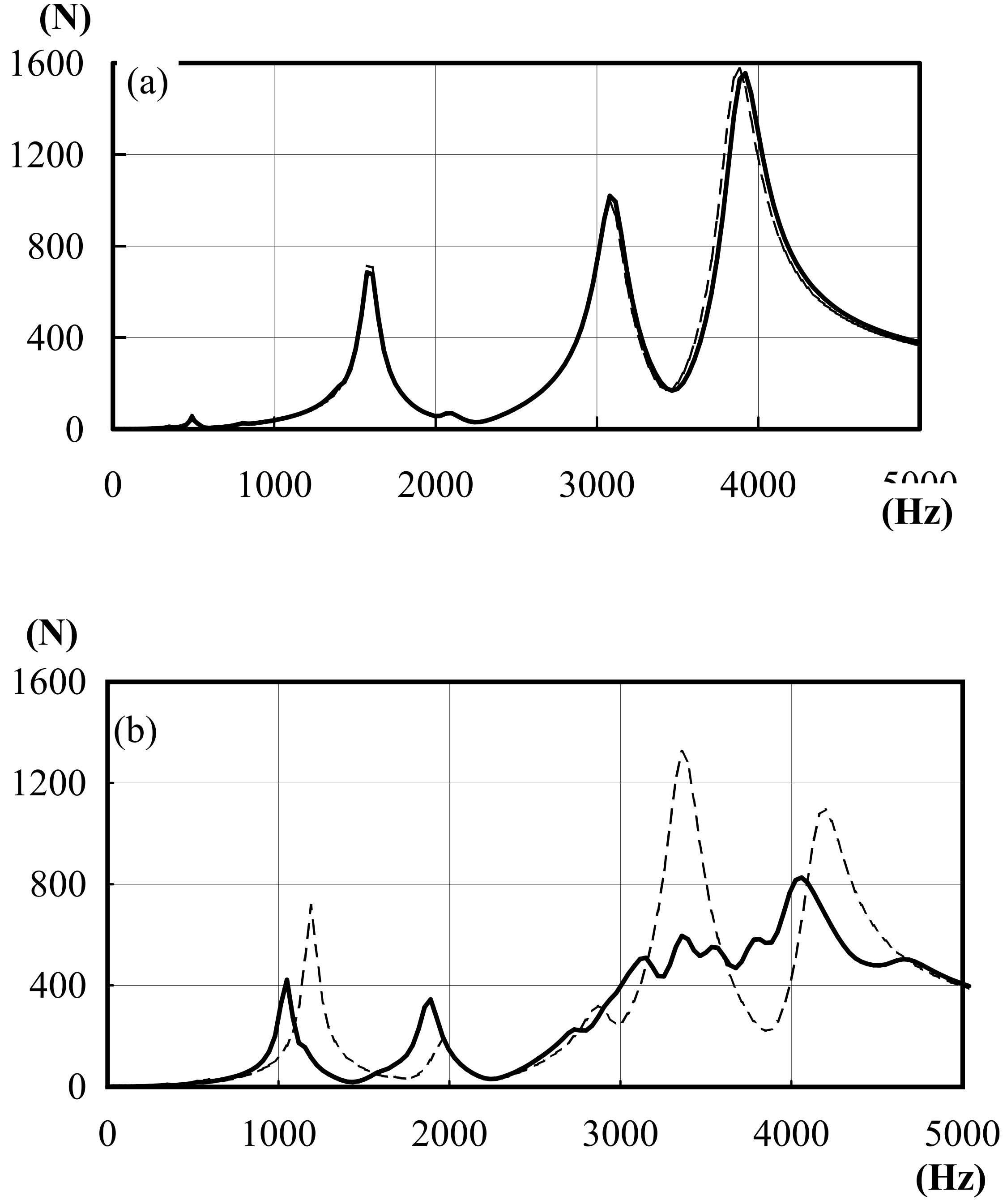

Figure 5. Evolution de la valeur efficace de l''effort de denture avec la fréquence d'engrènement. $K_{rot}=10^4$ Nm/rad, $K_{rad}=10^8$ N/m (a). $K_{rot}=10^6$ Nm/rad, $K_{rad}=10^9$ N/m (b).